\documentclass{elsart}
\usepackage{epsfig}
\usepackage{amssymb}
\usepackage{graphicx}
\usepackage{bm}
\newcommand{\BQ}{\begin{equation}}
\newcommand{\EQ}{\end{equation}}
\newcommand{\BQA}{\begin{eqnarray}}
\newcommand{\EQA}{\end{eqnarray}}
\newcommand{\be}{\begin{eqnarray}}
\newcommand{\ee}{\end{eqnarray}}

\newcommand{\del}{\partial}

\def\simge{\mathrel{%
   \rlap{\raise 0.511ex \hbox{$>$}}{\lower 0.511ex \hbox{$\sim$}}}}
\def\simle{\mathrel{
   \rlap{\raise 0.511ex \hbox{$<$}}{\lower 0.511ex \hbox{$\sim$}}}}

\begin{document}

\begin{flushright}
~\vspace{-1.25cm}\\
{\small\sf
 SPhT-T03/156\\ CPHT-RR-046.1003
}
\end{flushright}
\vspace{0.2cm}
\begin{frontmatter}

\title{Saturation and BFKL dynamics in the HERA data at small $x$}

\author{E.~Iancu,}
\author{K.~Itakura}
\address{Service de Physique Th\'eorique, CEA/DSM/SPhT, Unit\'e de recherche
associ\'ee au CNRS (URA D2306), CEA Saclay,
        F-91191 Gif-sur-Yvette, France}
\author{S.~Munier}
\address{Centre de Physique Th\'eorique,  Unit\'e mixte de recherche du CNRS (UMR 7644),
\'Ecole Polytechnique, F-91128 Palaiseau, France}
\date{\today}

\begin{abstract}
We show that the HERA data for the inclusive structure function 
$F_2(x,Q^2)$ for $x\le 10^{-2}$ and $0.045\le Q^2 \le 45\,{\rm GeV}^2$
can be well described within the color dipole picture, with a simple
analytic expression for the dipole-proton scattering amplitude,
which is an approximate solution to the non-linear evolution equations 
in QCD. For dipole sizes less than the inverse saturation momentum 
$1/Q_s(x)$, the scattering amplitude is the solution to the BFKL equation 
in the vicinity of the saturation line. It exhibits 
geometric scaling and scaling violations by the diffusion term. 
For dipole sizes  larger than $1/Q_s(x)$, the scattering amplitude 
saturates to one. The fit involves three parameters: 
the proton radius $R$, the value $x_0$ of $x$ at which the saturation 
scale $Q_s$ equals 1GeV, 
and the logarithmic derivative of the saturation momentum $\lambda$.
The value of $\lambda$ extracted from the fit 
turns out to be consistent with a recent calculation using the 
next-to-leading order BFKL formalism.
\end{abstract}
\end{frontmatter}
\vspace{1cm}

A main source of theoretical excitement about the high energy, or ``small--$x$'',
regime of deep inelastic lepton--hadron scattering (DIS), as currently investigated 
at HERA \cite{New_HERA}, is the possibility to reach a new regime of QCD 
\cite{GLR,MQ,MV,AM99,BKW,PI,SAT},
which is characterized by high parton densities, but remains within the realm
of perturbation theory, 
because the high density introduces a hard scale for the running of the QCD
coupling $\alpha_s$. For sufficiently high energies, perturbative
QCD consistently predicts that the small-$x$ gluons in a hadron wavefunction 
should form a Color Glass Condensate (CGC) \cite{PI}. This is a state characterized by
a hard saturation momentum $Q_s(x)$ which grows rapidly 
with $1/x$ \cite{GLR,AM99,EIM02,MT02,DT02}, and by large occupation numbers, 
of order $1/\alpha_s$, 
for all gluonic modes with transverse momentum less than or equal to $Q_s$
\cite{AM99,SAT}. The saturation momentum is a measure of the 
gluon density in the impact parameter space, and thus a
 natural scale for evaluating $\alpha_s$ at high energies.

Perturbative QCD also allows us to construct non-linear evolution 
equations \cite{BKW,PI} (see also Refs. \cite{GLR,MQ} for early versions
of these equations) which describe the formation and the properties
of the CGC, and its consequences for high energy scattering.
These equations resum all powers of $\alpha_s\ln(1/x)$, together
with the non-linear (higher-twist) effects responsible for gluon saturation
and for the unitarization of the scattering amplitude 
 at fixed impact parameter. So far, the general non-linear
equations are known only to leading order (LO) in $\alpha_s$, but the accuracy
is higher for the linear equation which governs the approach towards saturation
(say, when decreasing $x$ at fixed transverse resolution $Q^2 > Q_s^2(x)$) : This
is the BFKL equation \cite{BFKL}, which is presently known to next-to-leading
order (NLO) in $\alpha_s$ \cite{NLBFKL}, and whose renormalization-group improved version 
\cite{Salam99} has been recently used for a calculation of the energy dependence
of the saturation scale \cite{DT02}. The kinematical window for the 
validity of the BFKL equation in the presence of saturation has been estimated as 
$Q_s^2(x) < Q^2 < Q_s^4(x)/\Lambda_{\rm QCD}^2$ \cite{EIM02}.
Note that, with decreasing $x$, this  kinematical region
is pushed to higher  $Q^2$, where perturbation theory
is expected to work better.   

It is natural to ask whether this theoretical picture can be tested against
thef HERA data \cite{New_HERA}. The answer is a priori not obvious:
First, it is not clear whether the values of $x$ at HERA
are small enough for the corresponding saturation scale to be
much larger than $\Lambda_{\rm QCD}$. Second, the experimental points at HERA are
correlated in such a way that the smallest values of $x$ correspond also to rather
small values of $Q^2$ (of order $1\,{\rm GeV}^2$, or less), where the use of perturbation
theory becomes questionable. 

A first hint towards saturation effects at HERA came from 
the success of the simple ``saturation model'' by Golec-Biernat and 
W\"usthoff (GBW) \cite{GBW} which provided a reasonable description of the 
(old) HERA data \cite{Old_HERA} for $x \le  10^{-2}$ and all $Q^2$.
This model has been implemented within the color dipole picture 
which is also the framework for the present analysis. 
The dipole picture \cite{NZ91} is a factorization scheme for DIS,
 which is valid at small $x$ and is particularly convenient
for the inclusion of unitarity corrections. Specifically, the scattering
between the virtual photon $\gamma^*$ and the proton is seen as the 
dissociation of $\gamma^*$ into a  quark-antiquark pair (the ``color dipole'')
followed by the interaction of this dipole with the color fields in the 
proton. This leads to the following expressions for the $\gamma^* p$ cross-sections
and the $F_2$ structure function:
\vspace*{-0.3cm}
\be\label{sigmagamma}
F_2(x,Q^2)&=& \big(Q^2/4 \pi^2 \alpha_{\rm em}\big)\,\big(\sigma_T+\sigma_L\big),\nonumber\\
\sigma_{T,L}(x,Q^2)&=& \int dz d^2{\bm r}\, |\Psi_{T,L}(z,{\bm r},Q^2)|^2\,
\sigma_{\rm dipole}(x,{\bm r}).\,\,\,\,\ee
Here, $\Psi_{T,L}$ are light-cone wavefunctions for 
$\gamma^*$ (with transverse, or longitudinal, polarization), computable 
within QED (see, e.g., Ref. \cite{GBW} for explicit expressions to lowest order in
$\alpha_{\rm em}$).
Furthermore, $\sigma_{\rm dipole}(x,{\bm r})$ is the cross-section 
for dipole--proton scattering (for a dipole of transverse size ${\bm r}$),
and encodes all the information about hadronic interactions
(including unitarization effects). In Ref.  \cite{GBW}, the dipole
cross-section has been modelled as :
\vspace*{-0.5cm}
\be\label{Golec}
\sigma_{\rm dipole}(x,{\bm r})\,=\,\sigma_0\Big(1 - {\rm e}^{-{\bm r}^2 Q_s^2(x)/4}\Big)
\ee
where 
$Q_s(x)$ plays the role
of the saturation momentum, parametrized as $Q_s^2 (x)= (x_0/x)^\lambda$ GeV$^2$. 
Saturation is visible in the fact that the 
dipole scattering amplitude ${\mathcal N}(x,{\bm r})=1 - {\rm e}^{-{\bm r}^2 Q_s^2(x)/4}$
approaches the unitarity bound ${\mathcal N}=1$ for dipole sizes larger than
a characteristic size $1/Q_s(x)$, which decreases when decreasing $x$.
By fitting the three parameters $\sigma_0$, $x_0$ and $\lambda$, GBW managed
to give a rather good description of the old HERA data,
for both the inclusive and the diffractive structure functions.
More importantly --- since, in the mean time, the simple parametrization 
in eq.~(\ref{sigmagamma}) has been essentially ruled out \cite{BGBK} (see also below)
by the advent of a new set of data \cite{New_HERA}, with a much higher accuracy  ---, the 
original success of the ``saturation model'' allows us to draw a few interesting
lessons: 

{\it i\,})  The value of the saturation scale extracted from the fit is rather large
($Q_s^2 (x) >1$ GeV$^2$ for $x\le 10^{-4}$), which suggests that a perturbative 
approach, including more of the dynamics of QCD, may work rather well in the regime
where saturation effects become important.

{\it ii\,}) The inclusion of saturation, which tames the growth of $F_2$
with $1/x$ already at the hard scale $Q_s^2 (x)$, may cure the small--$x$ problem of the
linear evolution equations \cite{BFKL,DGLAP} 
without the need to resort to soft, non-perturbative, physics.

{\it iii\,}) The analysis of  Ref.  \cite{GBW} suggested a remarkable
regularity of the $F_2$ data, known as 
``geometric scaling'', which has been subsequently confirmed by
a model-independent analysis of the data \cite{geometric}: The measured 
total cross-section $\sigma_{\gamma^*p}(x,Q^2)=\sigma_T+\sigma_L$ 
shows approximate scaling as a function of the variable $Q^2/Q_s^2(x)$
with $Q_s^2(x) \propto 1/x^\lambda$ 
and $\lambda\sim 0.3$. This (approximate) scaling holds for small $x\le 10^{-2}$, 
and for all $Q^2  \le  450 \,{\rm GeV}^2$. 
Whereas at low momenta $Q^2 < Q_s^2(x)$ such a scaling
is indeed natural in the context of saturation, for larger 
$Q^2 > Q_s^2(x)$ it looks less natural, and it has been recently understood
\cite{EIM02,MT02,DT02} as a consequence of BFKL evolution with saturation
boundary conditions. 
The analysis in Refs. \cite{EIM02,MT02,DT02} predicts also
scaling {\it violations}, in particular via the BFKL diffusion term,
but so far these predictions have not been explicitly
tested against the HERA phenomenology.

These observations have motivated new attempts, better rooted in QCD, towards 
understanding the new HERA data \cite{New_HERA}
within the picture of saturation \cite{BGBK,KT,Levin}.
So far, these approaches have  mostly focused upon improving 
the behaviour of the fit at large $Q^2$,
by including DGLAP dynamics \cite{DGLAP}. 
Because of that, they led to rather successful global fits 
to the data (for $x\le 10^{-2}$ and all $Q^2$), 
but it is not clear to which extent such fits are sensitive to the
details of the  dynamics near, and at, saturation.
In Refs.~\cite{BGBK,KT}, the unitarization effects are included via the 
Glauber-like exponentiation of the leading-twist contribution, which is not
fully consistent with the non-linear dynamics in QCD. In Ref. \cite{Levin},
the correct non-linear dynamics {\it is} included (to LO in $\alpha_s$),
together with DGLAP dynamics, by a combined numerical analysis
of the  Balitsky-Kovchegov (BK) equation \cite{BKW} and of DGLAP equation. 
But also in that case, the success of the 
global fit turns out to be very much dependent upon the inclusion 
of DGLAP: Without the latter, that is, by using 
the BK equation {\it alone}, the fit of Ref. \cite{Levin} works satisfactory 
only for rather low $Q^2$ (up to a few GeV$^2$).
Besides, being fully numerical, the analysis in  Ref. \cite{Levin} does not allow
to simply test some qualitative features of BFKL dynamics towards saturation,
like geometric scaling and its violations.

Our aim in this Letter is to present a new analysis of the HERA data,
restricted to the kinematical range where one expects important
high density effects --- namely, $x\le 10^{-2}$ and $Q^2 < 50\,{\rm GeV}^2$ ---,
with the purpose of showing that these data are consistent with our present
understanding of BFKL evolution and saturation. The upper limit 
on $Q^2$ has been chosen large enough to include a significant number of ``perturbative''
data points, but low enough to justify the emphasis on BFKL, rather than
DGLAP, evolution (for the small-$x$ values of interest). 
In fact, $50\,{\rm GeV}^2$ is roughly consistent with the estimated upper bound 
$Q_s^4(x)/\Lambda_{\rm QCD}^2$ of the kinematical window for BFKL behaviour \cite{EIM02}, 
given our subsequent findings for $Q_s(x)$. Within this kinematical range,  we shall
provide a reasonable fit ($\chi^2/{\rm d.o.f.} \simeq 0.8-0.9$) to the new HERA data for 
$F_2$ based on a simple, analytic, formula for the dipole scattering amplitude, 
which is an approximate solution to the non-linear evolution equations in QCD \cite{BKW,PI}.
We shall refer to this fit as the ``CGC fit''.

It turns out that our fit involves the same set of free parameters 
(that is, $\sigma_0$, $x_0$ and $\lambda$) as the GBW ``saturation model''.
The need for the first two parameters, $\sigma_0$ and $x_0$, reflects the fact that,
even in a first principle calculation including saturation, some aspects of the
calculation remain non-perturbative: the impact parameter dependence of the
scattering amplitude and the initial condition at low energy.
Concerning the first aspect, we shall treat the proton as a homogeneous disk of radius $R$.
(A more realistic impact parameter dependence, which requires a model,
can be implemented along the lines of Refs. \cite{KT,Levin}. 
This is left for a later publication \cite{IIM}.)
Then, $\sigma_{\rm dipole}(x,{\bm r})=\sigma_0 {\mathcal N}(x,{\bm r})$ with 
$\sigma_0\equiv 2\pi R^2$ and ${\mathcal N}(x,{\bm r})$ given by the solution to the
homogeneous version of the non-linear evolution equation. Within the approximation 
that we shall use to solve this equation, the initial condition is fully
characterized by one parameter: the 
value $x_0$ of $x$ at which $Q_s$ equals 1GeV.
The third parameter, $\lambda$, which controls the energy dependence of the
saturation scale, is conceptually different, since this {\it can} be computed
in perturbation theory \cite{GLR,EIM02,MT02,DT02}. As already mentioned, this
parameter is presently known to NLO accuracy \cite{DT02}. More precisely,
the calculation in Ref. \cite{DT02} shows that $\lambda(Y) \equiv 
d\ln Q_s^2(Y)/dY$ (with $Y=\ln 1/x$) is not simply a constant, but rather
a slowly varying function, which decreases from $\lambda\approx 0.30$
for $Y=5$ to  $\lambda\approx 0.27$ for $Y=15$, with a theoretical uncertainty 
of about 15\%. Unfortunately, this uncertainty is still
too large to permit a good description of the data: $\lambda$ is the exponent
 which controls the growth of $F_2$ with $Y$, so the data
are very sensitive to its precise value. To cope with that, we shall
treat $\lambda$ as a free parameter, that we shall fit from the data. Remarkably,
the value of $\lambda$ that will come out from the fit, namely $\lambda \approx 0.25-0.29$,
is consistent with the theoretical prediction of Ref. \cite{DT02}, within the
theoretical uncertainty alluded to above\footnote{On the other hand, this value for
$\lambda$ is significantly smaller than that predicted by the LO BFKL equation with
running coupling \cite{GLR,EIM02,MT02,DT02}, which varies between $0.43$
and $0.37$ when increasing $Y$ from 5 to 10 \cite{DT02}. This discrepancy may explain why
the pure BK part of the fit in Ref. \cite{Levin} (based on a numerical solution 
to the BK equation \cite{BKW} with running coupling) is unable to accurately describe
the data with $Q^2 > 1\,\mbox{GeV}^2$ and small $x$.}.

The function ${\mathcal N}(Y,{\bm r})$
will be constructed by smoothly interpolating between two limiting 
behaviours which are analytically under control: the solution to the BFKL equation
for small dipole sizes, $r\ll 1/Q_s(x)$, and the Levin-Tuchin law \cite{LT99}
for larger dipoles, $r\gg 1/Q_s(x)$. As we shall check later, the quality of the fits 
is not very sensitive to the details of the interpolation, nor to the precise form
of the approach towards the limit  ${\mathcal N}=1$. On the other hand,
the data are quite sensitive to 
the details of the scattering amplitude at smaller sizes $r\le 1/Q_s(x)$, and thus provide
a test of BFKL dynamics. In this range, 
${\mathcal N}(Y,{\bm r})$ will be obtained via the saddle point approximation to
the LO BFKL equation, followed by an expansion to second order
around the saturation saddle point. As known from Refs. \cite{EIM02,MT02},
the first term in this expansion exhibits geometric scaling, while the second,
``diffusion'', term brings in scaling violations, which turn out to be crucial
for understanding the HERA data. The LO formalism is chosen for its simplicity:
it provides an explicit expression for ${\mathcal N}(Y,{\bm r})$, whose
physical interpretation is transparent. This is further motivated
by the observation \cite{DT02} that, with increasing $Y$,
the predictions of the (RG-improved) NLO BFKL formalism get closer and closer to those
of the LO equation with running coupling $\alpha_s(\mu^2)$, in which the scale $\mu^2$
for running is set either by the size of the dipole ($\mu^2=C/r^2$) \cite{MT02}, or
simply by the saturation momentum ($\mu^2=Q_s^2(x)$) \cite{EIM02}.
To account for the lack of accuracy of the LO formalism
at non-asymptotic $Y$, we shall treat, as announced, the saturation exponent $\lambda$ --- 
which is the critical parameter for describing the data --- as a free parameter.

Let us now briefly introduce our formulae. For both fixed coupling $\alpha_s$, or 
running coupling $\alpha_s(Q_s(Y))$, the solution to the
BFKL equation can be written in Mellin form as (for more details, see \cite{EIM02,MT02}):
\BQ
{\mathcal N}(Y,{\bm r})= \int_{C} \frac{d\gamma}{2\pi i} 
\left({\bm r}^2 Q_0^2\right)^{\gamma} {\rm e}^{h(Y) \chi(\gamma)}
\widetilde{\mathcal N}_0(\gamma) 
\EQ
\vspace*{-0.3cm}
where $Q_0$ is a reference scale of order $\Lambda_{\rm QCD}$,
$\chi(\gamma)=2\psi(1)-\psi(\gamma)-\psi(1-\gamma)$ with 
$\psi(\gamma)=d \ln \Gamma(\gamma)/d\gamma$,  
$\widetilde{\mathcal N}_0(\gamma)$ is the initial condition,
and the function $h(Y)$ depends upon our assumption on the running 
of $\alpha_s$: For fixed coupling, $h(Y)=\bar\alpha_s Y$ with 
$\bar\alpha_s=\alpha_s N_c/\pi$, while for a running coupling 
$\alpha_s(Q_s(Y))$, $h(Y)$ is determined by $dh/dY=\bar\alpha_s(Q_s(Y))$,
with $h(0)=0$.

In the saddle point approximation, valid when $h(Y)$ and
$\rho\equiv \ln (1/{\bm r}^2 Q_0^2)$ are large, 
\vspace*{-0.1cm}\BQ\label{SPN}
{\mathcal N}(Y,{\bm r})\simeq {\rm e}^{h(Y) F(\gamma_0({\mathcal R}), {\mathcal R})}
\EQ 
\vspace*{-0.2cm}
where ${\mathcal R}\equiv \rho/h(Y)$, $F(\gamma,{\mathcal R})=-\gamma {\mathcal R} + 
\chi(\gamma)$, and
$\gamma_0({\mathcal R})$ is the saddle point for a given ${\mathcal R}$, determined by:
\BQ\label{saddle}
\left. \frac{\del F}{\del \gamma}(\gamma,{\mathcal R})\right\vert_{\gamma=
\gamma_0({\mathcal R})}
= -{\mathcal R} + \chi'\big(\gamma_0({\mathcal R})\big)=0. 
\EQ
\vspace*{-0.4cm}
Note that, in the approximation above, we have neglected the initial condition
$\widetilde{\mathcal N}_0(\gamma)$, as well as the effect of the Gaussian fluctuations
around the saddle point: Indeed, these are slowly varying functions, which contribute
at most logarithmic terms (like $\ln h$ or $\ln \rho$) to the exponent in Eq.~(\ref{SPN}).

The saturation condition is written as: ${\mathcal N}(Y,{\bm r})={\mathcal N}_0$
for ${r} = 1/Q_s(Y)$, where ${\mathcal N}_0$ is a number of order one 
(its precise value is a matter of convention).
To the accuracy of interest, this condition implies
\BQ\label{sat_cond}
F(\gamma_0({\mathcal R}),{\mathcal R})\Big\vert_{{\mathcal R}={\mathcal R}_s}
\equiv \, -\,\gamma_0({\mathcal R}_s) {\mathcal R}_s +\chi\big(\gamma_0
({\mathcal R}_s)\big)=0,
\EQ
\vspace*{-0.3cm}
where ${\mathcal R}_s\equiv \rho_s(Y)/h(Y)$ and $\rho_s(Y) \equiv \ln (Q_s^2(Y)/ Q_0^2)$.

Eq.~(\ref{sat_cond}) shows that ${\mathcal R}_s$ is a pure number. Together with
Eq.~(\ref{saddle}), it allows us to compute both ${\mathcal R}_s$ and $\gamma_s\equiv 
\gamma_0({\mathcal R}_s)$
(the saddle point in the vicinity of the saturation line $Q^2=Q^2_s(Y)$).
One obtains  \cite{GLR,EIM02,MT02} : $\gamma_s = \chi(\gamma_s)/\chi'(\gamma_s)
=  0.627...$ and ${\mathcal R}_s = \chi'(\gamma_s)= 4.883....$
Note that  $\gamma_s$ is
not the same as the usual BFKL saddle point $\gamma_0=1/2$ (the latter would be obtained
by letting ${\mathcal R}\to 0$ in Eq.~(\ref{saddle})).
Whereas $\gamma_0$ describes the evolution with $Y$ at fixed, and relatively low, $Q^2$
(here, $Q^2\sim 1/r^2$), $\gamma_s$ corresponds rather to an evolution where,
when increasing $Y$, $Q^2$ is correspondingly increased, in such a way that the
condition $Q^2\sim Q_s^2(Y)$ (or $\rho \sim \rho_s(Y)$)  remains satisfied.

The dipole sizes ${\bm r}$ that we are interested in are smaller than,
but relatively close (in logarithmic units) to, $1/Q_s(Y)$. It is therefore
a very good approximation to compute ${\mathcal N}(Y,{\bm r})$ by expanding
the exponent in Eq.~(\ref{SPN}) to second order in $\rho-\rho_s =
\ln (1/{\bm r}^2 Q_s^2(Y))$. One then obtains \cite{EIM02,MT02}:
\vspace*{-0.5cm}
\BQA\label{NBFKL}
{\mathcal N}(Y,{\bm r})&\simeq & {\mathcal N}_0\,\exp\left\{ -\gamma_s (\rho-\rho_s)
-\frac{{\mathcal R}_s}{2\beta\rho_s}(\rho-\rho_s)^2\right\},
\EQA
where $\beta=\chi''(\gamma_s)= 48.518...$.
Eq.~(\ref{NBFKL}) shows that, in the vicinity of the saturation line ($\rho\approx \rho_s$),
where the quadratic term in the exponent can be neglected, the scattering amplitude 
scales as a function of ${\bm r}^2 Q_s^2(Y)$. This is geometric scaling.
The power $\gamma_s$ (or, more precisely, the difference $1-\gamma_s\approx 0.37$) 
is often referred to as an ``anomalous dimension''.
The  quadratic, ``diffusion'', term in Eq.~(\ref{NBFKL})
is responsible for scaling violations.

Now that we have found ${\mathcal R}_s$, one could use its original definition, namely
${\mathcal R}_s=(1/h(Y))\ln (Q_s^2(Y)/ Q_0^2)$, together with 
the equation $dh/dY=\bar\alpha_s(Q_s(Y))$ to compute $Q_s(Y)$ \cite{EIM02,MT02}.
However, if one does so,
then the resulting expression for $Q_s(Y)$ will not be accurate enough
to describe the data. Rather, we shall rely on the NLO calculation in Ref. \cite{DT02} 
to conjecture that $\rho_s(Y) =\lambda Y$ 
with $\lambda$ a pure number, to be fitted from the data. As for the other coefficients
which appear in Eq.~(\ref{NBFKL}), namely $\gamma_s$ and the ratio
 $\kappa\equiv \beta/{\mathcal R}_s$,
these will be kept as in the LO BFKL approximation since ({\it a}\,) their
numerical values do not change appreciably when going to NLO
\cite{DT02}, and ({\it b}\,) the fit is not very sensitive to their precise
 values, as we shall check. 
This last feature --- the stability of the fit with respect to small changes
in the values of the  coefficients  $\gamma_s$ and $\kappa$ --- is important since,
given the present status of the theory, it seems to be quite difficult to improve
our theoretical control of these parameters (e.g., by going to NLO accuracy). 
The solution ${\mathcal N}(Y,{\bm r})$ to the (RG-improved) NLO BFKL equation with saturation
boundary condition is known \cite{DT02} only for asymptotically large
$Y$, and thus cannot be applied to the kinematical range at HERA without introducing
additional free parameters. Therefore, using that (rather complicated) solution in a fit
would obscure the physics without automatically improving the predictive power of the theory.

To summarize, the dipole cross-section that we shall use in the CGC fit reads 
$\sigma_{\rm dipole}(x,{\bm r})=2\pi R^2 {\mathcal N}(rQ_s,Y)$, with
\vspace*{-0.2cm}
\be\label{NFIT}
{\mathcal N}(rQ_s,Y)
&=&{\mathcal N}_0\, \left(\frac{{r} Q_s}{2}\right)^
{2\big(\gamma_s + \frac{\ln(2/rQ_s)}{\kappa \lambda Y}\big)}\,\,\quad{\rm for}\quad rQ_s\le 2,
\nonumber\\
{\mathcal N}(rQ_s,Y)&=& 1 - {\rm e}^{-a\ln^2(b\, rQ_s)}\qquad\quad\qquad{\rm for}\quad rQ_s > 2,
\ee
where $Y=\ln(1/x)$,
$Q_s\equiv Q_s(x) = (x_0/x)^{\lambda/2}$ GeV,
and we have redefined $Q_s$ in such a way that ${\mathcal N}(rQ_s,Y)={\mathcal N}_0$
for $rQ_s=2$ (to facilitate the comparison with the GBW results based on
Eq.~(\ref{Golec})). The expression in the second line has the correct functional
form for $r\gg 2/Q_s$, as obtained either by solving the BK equation \cite{LT99}, 
or from the theory of the CGC \cite{SAT} (see also \cite{IM03} for a more careful
discussion). This is strictly valid only to LO accuracy,
but here it is used merely as a convenient interpolation 
towards the `black disk' limit ${\mathcal N}=1$.
(The details of this interpolation 
are unimportant for the calculation of $\sigma_{\gamma^*p}$.)
The coefficients $a$ and $b$ are determined  uniquely
from the condition that ${\mathcal N}(rQ_s,Y)$ 
and its slope be continuous at $rQ_s=2$.
The overall factor ${\mathcal N}_0$ in the first line of Eq.~(\ref{NFIT})
is ambiguous, reflecting an ambiguity in the definition of $Q_s$. 
We shall repeat our fit for various values
of ${\mathcal N}_0$ between $0.5$ and $0.9$ and check that the results of the fit
change only little. 

As already mentioned, the coefficients $\gamma_s$ and 
$\kappa$ are fixed to their LO BFKL values: $\gamma_s= 0.63$ and $\kappa= 9.9$.
We shall use the same photon wavefunctions $\Psi_{T,L}$ as in Refs. \cite{GBW,BGBK,KT}.
These involve a sum over the active quark flavors, and in our fits
we shall work with three quarks of equal
mass $m_q$, for which we shall choose three different values:
$m_q=140$ MeV,  $50$ MeV, and $10$ MeV.
(Further analysis of the $m_q$--dependence of the fit, as well as the inclusion
of the heavier, charm quark are left for Ref. \cite{IIM}.)
Thus, the only free parameters of the fit are $R$, $x_0$ and $\lambda$,
as announced.

Our fit has been performed for the $F_2$ data at ZEUS 
(the first two references in \cite{New_HERA}) with $x\leq 10^{-2}$ and $Q^2$
 between 0.045 and 45 $\mbox{GeV}^2$ 
(156 data points). Only ZEUS data have been considered because 
there is a mismatch between H1 and ZEUS concerning the data normalization,
and it is only ZEUS which has data
at very low $Q^2$, in the saturation region. (Adding the H1 data in the
fit would have implied a normalization adjustment, and would have put
more weight to the moderate and large $Q^2$ data.)
Statistical and systematic errors have been added in quadrature.

 The values obtained for the various parameters, and the $\chi^2/{\rm d.o.f.}$ for the
fits corresponding to different choices for ${\mathcal N}_0$ and $m_q$,
are shown in Tables 1 and 2. 
Note, in particular, the good stability of the result for $\lambda$,
which changes only by 15\% (within the range $\lambda = 0.25-0.29$)
when varying ${\mathcal N}_0$ and $m_q$. As anticipated, this value of $\lambda$
is in agreement with the theoretical calculation at NLO in Ref. \cite{DT02}.
In Table 1, we also show, for comparison,
the results obtained when performing a fit to the same set of data with the
GBW cross-section (\ref{Golec}). Clearly, the corresponding  $\chi^2/{\rm d.o.f.}=1.59$ 
is relatively poor. Note also that the value of $Q_s^2(x)$ emerging from the CGC
fit is smaller --- roughly, by a factor of $2$ for ${\mathcal N}_0=0.7$
--- than the corresponding value for the GBW fit (since $x_0$ is correspondingly smaller).

In Figs. 1 and 2, the results of the fit are plotted against the data
 for ${\mathcal N}_0=0.7$ and $m_q=140$ MeV. In Fig. 2, we have also shown the
extrapolation of our fit towards larger values of $x$ and $Q^2$ (outside the range of
the fit), together with the corresponding HERA data, in order
to emphasize that the deviations from BFKL dynamics --- due notably to 
the presence of the valence quarks and to the DGLAP
evolution \cite{DGLAP}  ---  do eventually show up, as expected.

In both Figs. 1 and 2, we also show (with dashed line)
the prediction of the BFKL calculation without saturation, as obtained
by extrapolating the formula\footnote{Note that this is not the standard BFKL
formula, since the saddle point $\gamma_s$ is different from $\gamma_0=1/2$.
But the behaviour illustrated by the dashed line fit in  Figs. 1 and 2 is 
representative for the BFKL fits without saturation  \cite{BFKL_fits}.
For instance, a similar  behaviour can be found by extrapolating the fit
in the last paper of Ref. \cite{BFKL_fits} towards low $Q^2$.}
in the first line of Eq.~(\ref{NFIT}) to arbitrarily 
large $rQ_s$ (with the same values for the parameters as in Table 1, and the
infrared diffusion term switched off for $rQ_s > 2$). 
This pure BFKL fit shows a too strong increase with $1/x$ at small $Q^2$, 
as expected from similar analyses in the literature \cite{BFKL_fits}.
On the other hand, the complete fit, including saturation, works remarkably well
down to the lowest values of $Q^2$ that we have included. Since, a priori, such low
values of $Q^2$ seem to be completely out of the reach of perturbation theory, it is
important to explain why, in the context of saturation, it is still meaningful to
approach these data via the previous calculation:

\bigskip
\begin{table}[ht]
{
\begin{center}
\begin{tabular}{l||c|c|c|c|c|c|c||c}
${\mathcal N}_0$/model                 
  & $0.5$ 
  & 0.6
  & 0.7 
  & 0.8 
  & 0.9
  & GBW \\
\hline
$\chi^2$
  & 146.43
  & 129.88
  & 123.63
  & 125.61
  & 133.73
  & 243.87\\
$\chi^2/\mbox{d.o.f}$ 
  & 0.96
  & 0.85
  & 0.81
  & 0.82
  & 0.87
  & 1.59\\
\hline
$x_0\ (\times 10^{-4})$   
  & 0.669
  & 0.435
  & 0.267
  & 0.171
  & 0.108
  & 4.45\\
$\lambda$
  & 0.252
  & 0.254
  & 0.253
  & 0.252
  & 0.250
  & 0.286\\
$R$ (fm) 
  & 0.692
  & 0.660
  & 0.641
  & 0.627
  & 0.618
  & 0.585\\
\end{tabular}
\end{center}
}
\bigskip
\caption{The CGC fits for different values of ${\mathcal N}_0$
and 3 quark flavors with mass $m_q = 140$ MeV. Also shown is
the fit obtained by using the GBW model, Eq.~(\ref{Golec}).}
\end{table}
\begin{table}[ht]
{
\begin{center}
\begin{tabular}{l||c|c|c||c|c|c|}
 & \multicolumn{3}{|c||}{$m_q=50\ {\rm MeV}$}
 & \multicolumn{3}{|c|}{$m_q=10\ {\rm MeV}$}\\
\hline
${\mathcal N}_0$
  & 0.5 
  & 0.7 
  & 0.9 
  & 0.5
  & 0.7 
  & 0.9 \\
\hline
$\chi^2$
  & 148.02
  & 108.52
  & 108.76
  & 149.27
  & 107.64 
  & 106.49\\
$\chi^2/\mbox{d.o.f}$ 
  & 0.97
  & 0.71
  & 0.71
  & 0.98
  & 0.70
  & 0.70 \\
\hline
$x_0\ (\times 10^{-4})$   
  & 2.77
  & 0.898
  & 0.333
  & 3.32
  & 1.06 
  & 0.382\\
$\lambda$
  & 0.290
  & 0.281
  & 0.274
  & 0.295
  & 0.285
  & 0.276\\
$R$ (fm) 
  & 0.604
  & 0.574
  & 0.561
  & 0.593
  & 0.566
  & 0.554 \\
\end{tabular}
\end{center}
}
\bigskip
\caption{The CGC fits for three values of ${\mathcal N}_0$ and quark masses $m_q = 50$ MeV
  (left) and $m_q = 10$ MeV (right).}
\end{table}

\begin{figure}
  \centerline{
  \epsfsize=1.\textwidth
\epsfbox{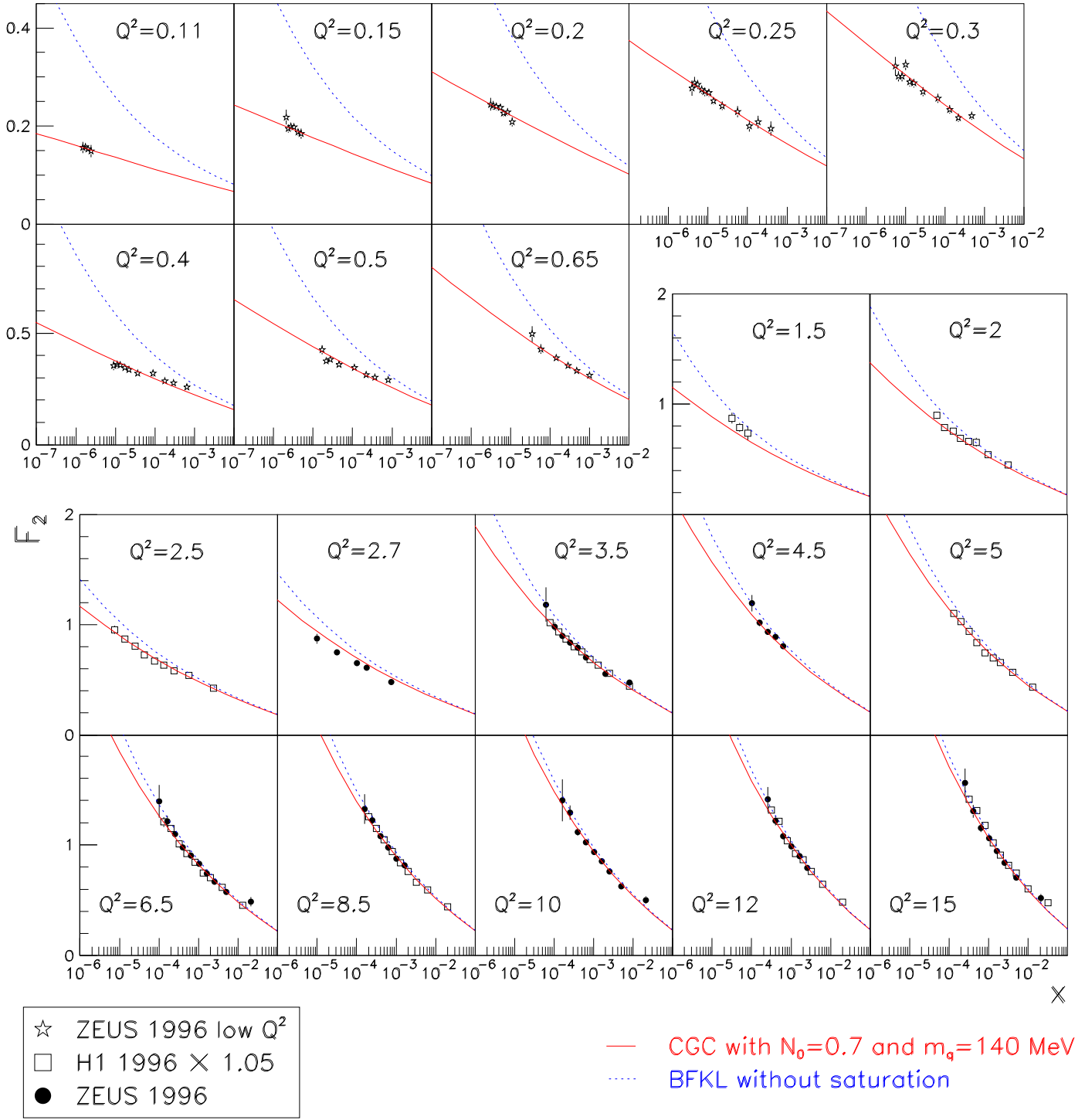}
  }
\vspace*{-.5cm}
 \caption[]{The $F_2$ structure function in bins of $Q^2$ for 
small (upper part) and moderate (lower part) values of $Q^2$.
The experimental points are the latest published
data from the H1 and ZEUS collaborations \cite{New_HERA}.
(The H1 data have been rescaled by a factor $1.05$ which is within
the normalization uncertainty.)
The few data points at the lowest available $Q^2$ 
(0.045, 0.065 and $0.085\
\mbox{GeV}^2$) are not displayed although they are included in the fit.
The full line shows the result of the CGC fit with ${\mathcal N}_0=0.7$ to the ZEUS data
for  $x\leq 10^{-2}$ and $Q^2\leq 45\ \mbox{GeV}^2$. 
The dashed line shows
the predictions of the pure BFKL part of the fit (no saturation).}
\end{figure}
\begin{figure}
  \centerline{
  \epsfsize=1.\textwidth
\epsfbox{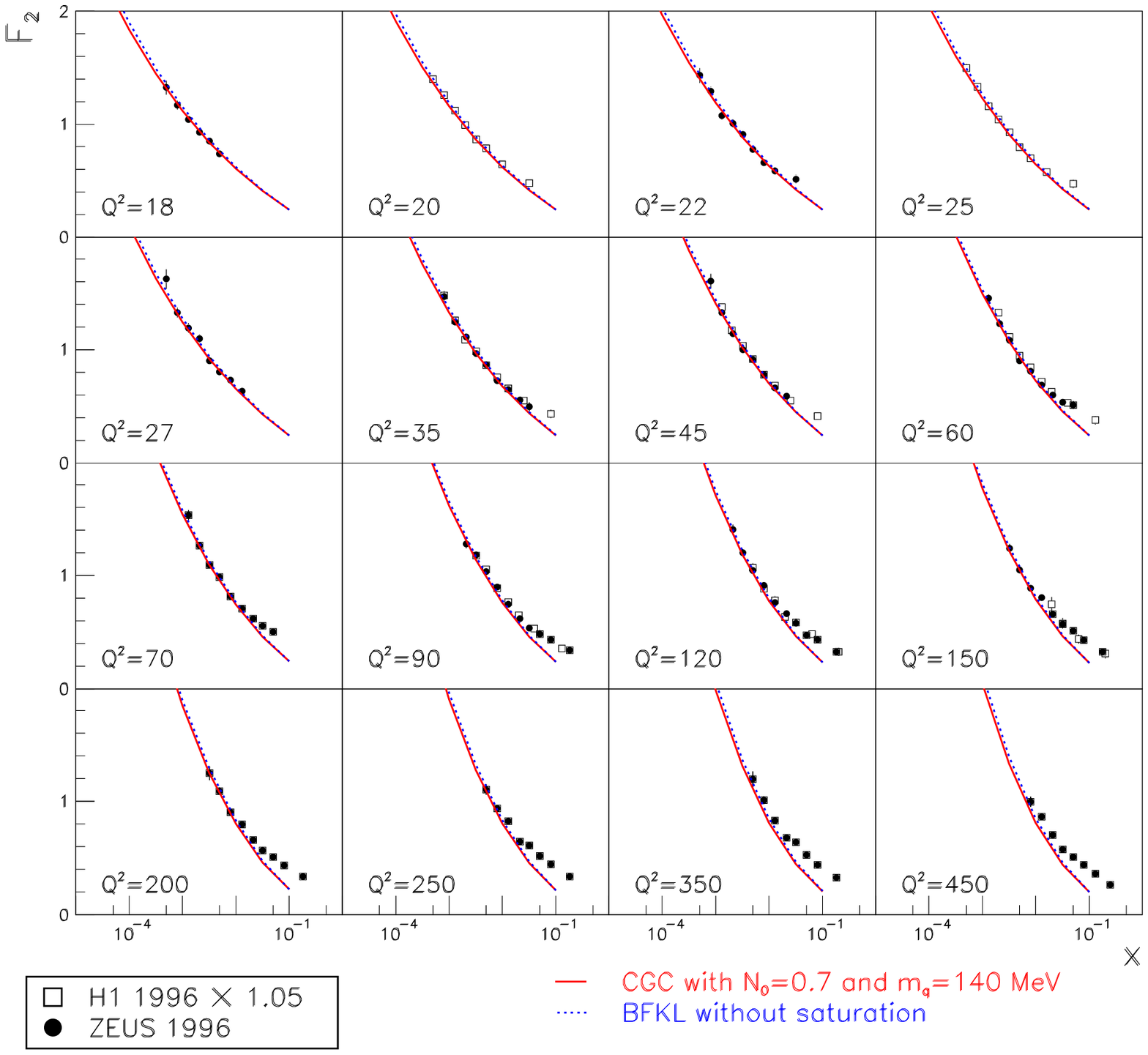}
  }
 \caption[]{The same as in Fig. 1, but for large $Q^2$. 
Note that in  the bins with $Q^2\ge 60 \,{\rm GeV}^2$, the CGC fit 
is extrapolated outside the
range of the fit ($Q^2<50\ \mbox{GeV}^2$ and $x\leq 10^{-2}$), to better emphasize
its limitations.}
\end{figure}

The reason is that, for the low $Q^2$ data at HERA, the associated
values of $x$ are also very small, so that the corresponding
saturation scale is relatively large:
$Q_s^2(x)\sim 1.3-2.3 \,{\rm GeV}^2$ for $x\sim 10^{-5}-10^{-6}$.
Thus, for these data, we have $Q^2\ll Q_s^2(x)$, which implies that the large,
non-perturbative dipoles ($r\simge 1/Q$) are deeply in the saturation regime,
so for them ${\mathcal N}\approx 1$ independently of the detailed mechanism
leading to such a strong absorbtion. Clearly, the $q\bar q$ pairs with transverse
sizes $r> 2\,\mbox{GeV}^{-1}$ have a strong overlap with hadronic states,
like the $\rho$ meson, whose correct treatment goes beyond the scope of perturbation theory.
But in the calculation of the DIS cross-section $\sigma_{\gamma^*p}$ at small $x$,
all such states are summed up with the same weight
(since they all have an amplitude ${\mathcal N}\sim 1$), so this sum 
simply accounts for the total probability for hadronic components with 
$Q^2 < 1 \,\mbox{GeV}^2$ in the virtual photon wavefunction. 
It is reasonable to assume that this total probability is correctly
given by perturbation theory (i.e., by summing over all $q\bar q$ 
states with sizes $r> 2\,\mbox{GeV}^{-1}$), although the way how this probability 
is distributed among various meson states is truly non-perturbative\footnote{We would like
to thank Al Mueller for pointing out this argument to us.}.
Note also that, because of saturation, the distribution of the  dipole
sizes in the integrand of Eq.~(\ref{sigmagamma}) when $Q^2\ll Q_s^2(x)$ is pushed
towards smaller dipoles, at least to logarithmic accuracy \cite{GBW,KT,IIM} :
the typical dipole sizes which contribute to the 
integration are logarithmically distributed in the range $Q^2\ll 1/r^2\ll Q_s^2(x)$.

As anticipated, we have checked that neither the quality of the fit, nor the value of
the parameters, change appreciably when the coefficients $\gamma_s$ and 
$\kappa$ in Eq.~(\ref{NFIT}) are modified by about $10\,\%$ (but the fit
appears to be more sensitive\footnote{One should also keep in mind that these two parameters
are actually correlated, so it makes no sense to consider strong, {\it independent}, 
variations in their values. For instance, in the LO BFKL formalism that we are using
here, they are both determined by properties of the function $ \chi(\gamma)$ near the
saturation saddle point $\gamma_s$.} 
to the value of $\gamma_s$ then to that of $\kappa$) \cite{IIM}. Such variances 
are representative for the theoretical uncertainties of our BFKL calculation,
so it is reassuring to see that the success of our fit is not
dependent upon a fine-tunning of these parameters.

 Finally,
in order to study the role of scaling violations, we have performed a fit with
a pure scaling function \cite{IIM03}, namely: ${\mathcal N}(rQ_s) = {\mathcal N}_0
 (rQ_s/2)^{2\gamma}$ (with $\gamma$
a free parameter) for $rQ_s \le 2$, and saturation (as in the second line of
Eq.~(\ref{NFIT})) for $rQ_s > 2$ (see results in Table 3). 
We have found that, although this fit involves 4 parameters, its quality
is relatively poor: $\chi^2/{\rm d.o.f.}\approx 1.4$. More interestingly,
we have found that the ``best'' value of $\gamma$ for describing the data
in the kinematical range of interest (and within this scaling Ansatz) 
is $\gamma \approx 0.84$, which is significantly smaller than one 
(at variance with the GBW ``saturation model'', and also with DGLAP 
evolution, which would predict a power-law behaviour with $\gamma =1$ 
up to logarithmic corrections), 
but also significantly larger than the LO BFKL value $\gamma_s=0.63$.
This clearly shows that the presence of the diffusion term in Eq.~(\ref{NBFKL})
has been essential for the success of the 3-parameter CGC fit: This term
 violates scaling for $r < 2/Q_s$, and  enhances the effective 
``anomalous dimension'' 
from $\gamma_s$ to 
\vspace*{-0.2cm}\be
\gamma_{\rm eff}(rQ_s,Y)\,\equiv\,-\,\frac{d \ln {\mathcal N}(rQ_s,Y)}{d\ln(4/r^2Q_s^2)}\,=\,
\gamma_s + 2\,\frac{\ln(2/rQ_s)}{\kappa \lambda Y}\,.
\ee
The difference $\gamma_{\rm eff}(rQ_s,Y) - \gamma_s$
decreases with $Y$, but increases with the deviation
$\rho-\rho_s= 2\ln(2/rQ_s)$ from the saturation line. This behaviour
is illustrated in Fig. 3, where we have also included, for comparison,
the pure scaling functions with $\gamma_s=0.63$ and
$\gamma = 0.84$ (the average ``anomalous dimension'' preferred by the fit in Table 3).
Note that, for any $x$, there is a significant range, roughly at $1 < rQ_s < 2$, 
within which the geometric scaling approximation with $\gamma_s=0.63$ works quite well.
On the other hand, for $rQ_s \ll 1$,  $\gamma_{\rm eff}$ gets closer to
one\footnote{Taken literally, our formula implies
$\gamma_{\rm eff}\to \infty$ as $r\to 0$,
which together with Eq.~(\ref{NFIT}) implies that, when $r\to 0$, 
the dipole amplitude vanishes faster than any power of $r$. However, this 
(unphysical) behaviour
has no influence on the fit, since in the kinematical range of interest the
integrated result for $\sigma_{\gamma^*p}$ is almost insensitive to very 
small dipoles.}, and thus mimics DGLAP behaviour; 
within the range of the fit, this is important only
for the data with relatively large $Q^2$ and not so small values of $x$, which
are more sensitive to such small-size dipoles.

\bigskip
\begin{table}[ht]
{
\begin{center}
\begin{tabular}{l||c|c||c|c|c|c|c|}
${\mathcal N}_0$
& $\chi^2$
& $\chi^2/\mbox{d.o.f}$ 
& $x_0\ (\times 10^{-4})$   
& $\lambda$
& $R$ (fm) 
& $\gamma$\\
\hline
  0.7
  & 215.70
  & 1.42
  & 3.79
  & 0.313
  & 0.572
  & 0.845\\
\end{tabular}
\end{center}
}
\bigskip\caption{A 4 parameter fit for geometric scaling + saturation.}
\end{table}

Further variations and checks of our fits, together with more applications
to the HERA phenomenology, will be presented somewhere else \cite{IIM}.

\begin{figure}
  \centerline{
  \epsfsize=1.\textwidth
\epsfbox{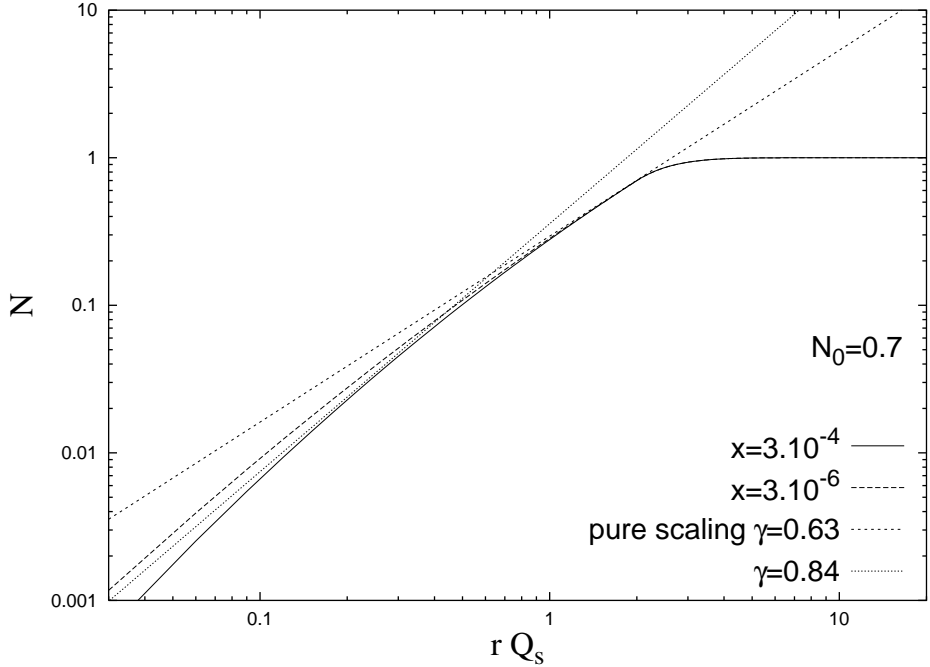}
  }
 \caption[]{The dipole amplitude for two values of $x$, compared to the pure 
scaling functions with ``anomalous dimension'' $\gamma=\gamma_s=0.63$ and $\gamma=0.84$.}
\end{figure}

\vspace*{-0.8cm}
\section*{Acknowledgments}
\vspace*{-0.6cm}
We would like to thank Fran\c cois Gelis,
Larry McLerran, Al Mueller, and Dionysis Triantafyllopoulos
for carefully reading the manuscript and many useful observations. Also,
fruitful conversations with Ian Balitsky, Stefano Forte, Gavin Salam and, especially,
Robi Peschanski are gratefully acknowledged.


\end{document}